\documentstyle[epsfig]{aipproc}
\begin{document}
\title{
Study of the $\eta\pi^o$ system
           in the mass range up to 1200 MeV}
\author{GAMS Collaboration}
\address{
D.Alde, F.G.Binon, M.Boutemeur, C.Bricman,
         S.V.Donskov, M.Gouanere,\\ A.V.Inyakin, ~S.Inaba, V.A.Kachanov,
         G.V.Khaustov, E.A.Knapp, ~A.A.Kondashov, A.A.Lednev,
         V.A.Lishin, J.P.Peigneux, M.Poulet,~\framebox{Yu.D.Prokoshkin}~,
         S.A.Sadovsky, V.D.Samoylenko, P.M.Shagin, A.V.Singovsky,
         A.E.Sobol, J.P.Stroot, V.P.Sugonyaev, K.Takamatsu, T.Tsuru}
\author{
                     {\it presented by} ~S.A.Sadovsky}
\address{
\vspace*{-0.45cm}
               IHEP,~Protvino,~Moscow~reg.,~142284,~RUSSIA
\vspace*{-0.1cm}
}
\maketitle

\begin{abstract}
The reaction $\pi^-p \rightarrow \eta\pi^o n$ has been studied
with GAMS-2000 spectrometer in the secondary 38 GeV/c $\pi^-$-beam 
of the IHEP U-70 accelerator. Partial wave analysis of the
reaction has been performed in the $\eta\pi^o$ mass range up to
1200 MeV. The $a_0(980)$-meson is seen as a sharp peak in S-wave.
The $t$-dependence of $a_0(980)$ production cross section has been 
studied. Dominant production of the $a_0(980)$ at a small transfer 
momentum $t$ confirms the hypothesis of Achasov and Shestakov about 
significant contribution of the $\rho_2$ exchange ($I^GJ^{PC}=1^+2^{--}$) 
in the mechanism of $a_0(980)$ meson production in $t$-channel of
the reaction.
\end{abstract}

The $a_0(980)$-meson is quite peculiar object in meson spectroscopy.
More than two decades of extensive experimental and theoretical studies
of the $a_0$ meson have been undertaken. Nevertheless its nature is still 
not completely clear, see for example \cite{a0}.

In the present talk the results on the $a_0(980)$-meson production
in the charge exchange reaction
\begin{equation}
\pi^-p \rightarrow \eta\pi^o n,
\end{equation}
at 38 GeV/c are presented for $4\gamma$ final states of the
$\eta\pi^o$-system. The data were collected with the GAMS-2000
multiphoton spectrometer in the secondary $\pi^-$-beam of the U-70
accelerator of IHEP. In total $140 \times 10^3$ events of reaction
(1) were collected. Further details of the experiment as well as
\begin{figure*}[thb]
\centerline{\epsfig{file=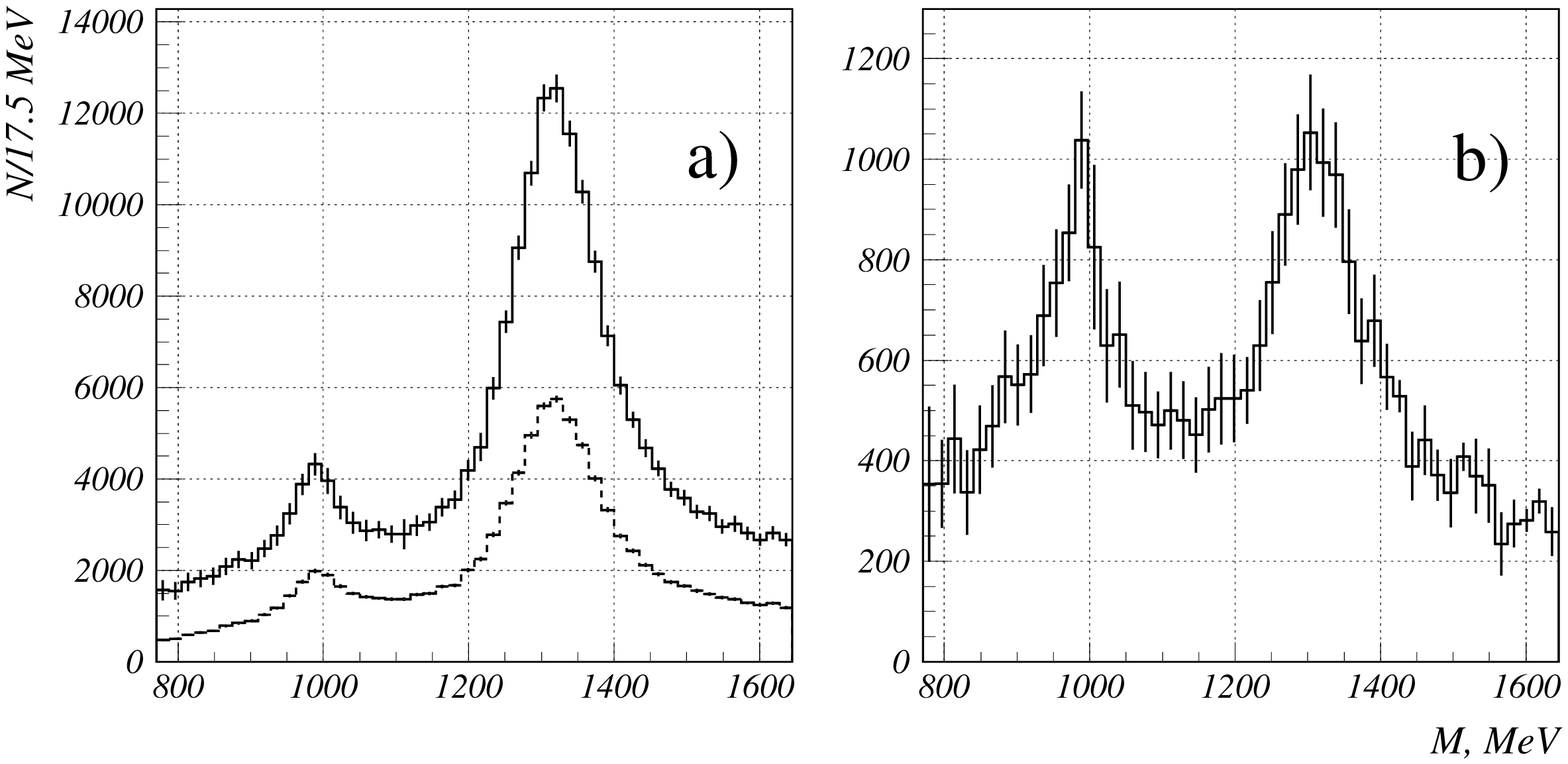,height=2.3in,width=4.6in}}
\vspace{5pt}
\caption{Spectra mass of $\eta\pi^o$ system in reaction (1): a)
the measured spectrum at $-t< 1~~(GeV/c)^2$, dashed line;
the efficiency corrected spectrum, solid line; b) the efficiency
corrected spectrum at $-t< 0.05~~(GeV/c)^2$.}
\end{figure*}
the data treatment procedures can be found elsewhere \cite{etapi,a2}.

The mass spectra of the $\eta\pi^o$-systems produced in reaction
(1) at \hbox{$-t < 1 ~(GeV/c)^2$} are shown in Fig.~1a. The efficiency
corrected spectrum at \hbox{$-t < 0.05 ~(GeV/c)^2$} is shown in Fig.~1b.
The efficiency was calculated by Monte-Carlo method, details see 
elsewhere \cite{eff}. Two peaks, corresponding to the $a_0(980)$
and $a_2(1310)$ mesons, are clearly seen in both figures, 
but their intensities are different. While at  $-t < 1 ~(GeV/c)^2$, 
see Fig.~1a, solid line, the $a_2(1310)$-meson dominates in the 
spectrum, at $-t < 0.05 ~(GeV/c)^2$, the intensities of both peaks are 
actually similar, Fig.~1b, i.e. the production mechanisms of these mesons 
in reaction (1) are different.

In the $a_2(1310)$ meson production the natural spin-parity exchanges
(mainly $\rho$-exchange) dominate in the $t$-channel of the reaction \cite{a2},
that leads to a suppression of the $a_2$ production at $t \sim 0$
\cite{Tuan-Achasov}. As for the $a_0(980)$ meson production the 
only unnatural exchanges are allowed in the $t$-channel. Moreover, 
due to the features of reaction (1) one should expect a significant
contribution of the $\rho_2$ exchange ($I^GJ^{PC}=1^+2^{--}$) usually
hidden in other reactions, see \cite{Achasov2}. The last exchange leads to 
a non-vanishing differential cross section of the $a_0(980)$ production in
reaction (1) at small $t$. Therefore, a study of the $a_0(980)$ differential 
cross section is also important for understanding of the status of the 
$\rho_2$ states.

For selection of the $a_0(980)$ events a Partial Wave Analysis
(PWA) of the $\eta\pi^o$ system produced in reaction (1) has been
performed in 17.5 $MeV$ mass bins in the range up to 1200 MeV taking
into account $S$, $P_0$, $P_-$ and $P_+$ waves. The PWA procedure as
well as a solution of the ambiguity problem are described elsewhere
\cite{s8}. Here it would be useful to mention that in the PWA
model with $S$ and $P$ waves there are only two non-trivial solutions.
Both solutions at $t < 1~(GeV/c)^2$ are presented in Fig.~2. The
physical solution can be identified as that one with the resonance 
peak in $S$-wave, solid line in Fig.~2.
The parameters of the $a_0(980)$ resonance
\begin{equation}
          M = 992 \pm 3~~MeV, ~~~~~~~~~~~\Gamma =  90 \pm 9~~MeV
\end{equation}
\begin{figure*}[tbh]
\centerline{\epsfig{file=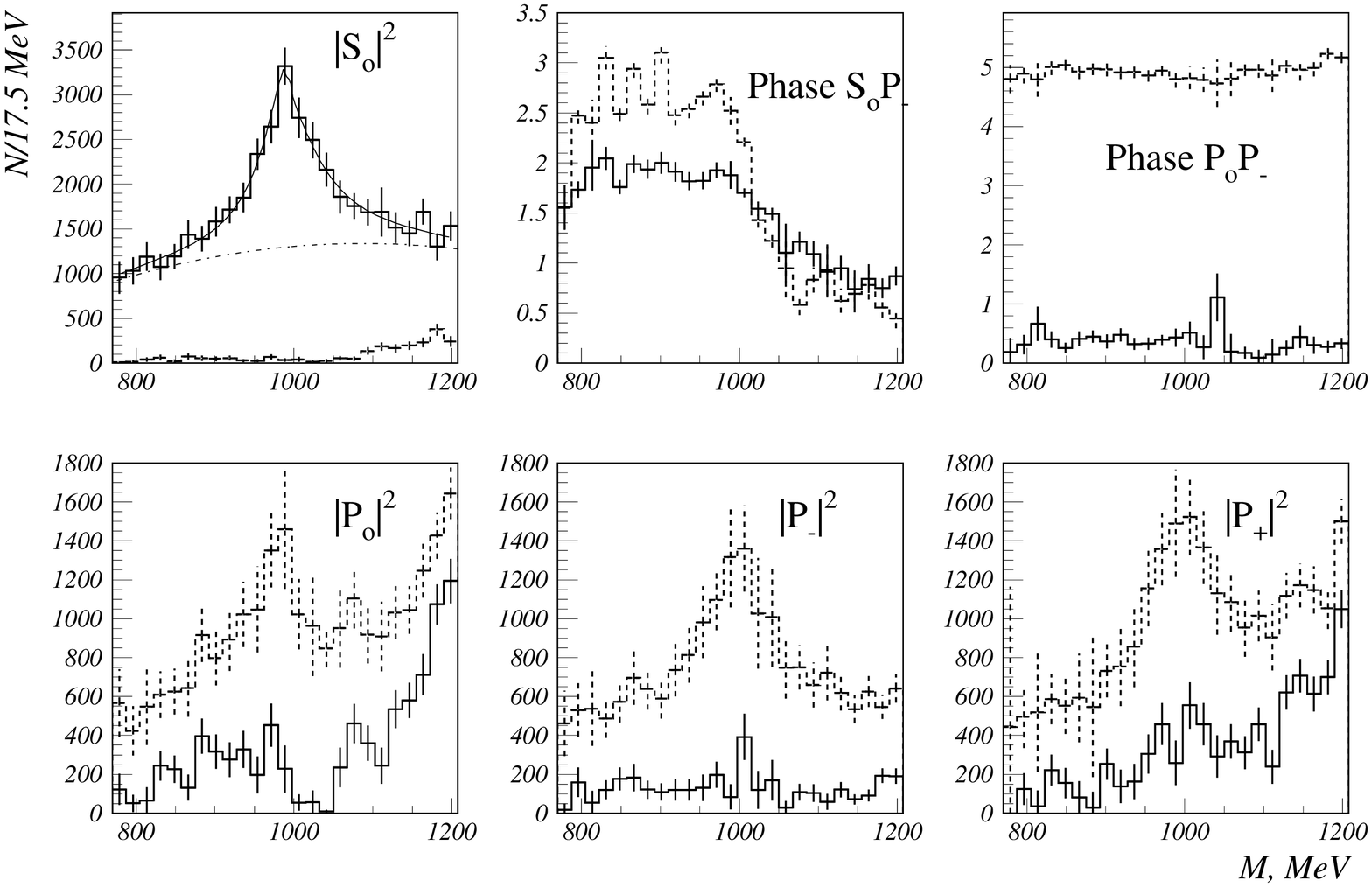,height=3in,width=5in}}
\vspace{5pt}
\caption{ Two non-trivial solution in the PWA of reaction (1) in
the case of $S$, $P_0$, $P_-$ and $P_+$ waves. The physical 
(unphysical) solution is shown by solid (dashed) line.}
\end{figure*}
were obtained by fitting the $S$-wave intensity in the physical
solution to the uncoherent sum of Breit-Wigner function and polynomial
background, 
Fig.~2. Actually the same resonance parameters were obtained if 
Flatte's \cite{Flate} or Achasov's \cite{Achasov3} formulae
where used for parameterization of the $a_0(980)$ resonance.
The number of events in the $a_0(980)$ peak normalized for the
cross section of the $a_2(1320)$ production in $D_+$ wave of
reaction (1) was used for the cross section determination of
the $a_0(980)$-meson, see \cite{a4}:
\begin{equation}
 \sigma(\pi^-p \rightarrow a_0 n)\times BR(a_0 \rightarrow \eta\pi^o)
                                                          = 68 \pm 25 ~~nb.
\end{equation}

To obtain the $t$-dependence of the $a_0(980)$ production in
reaction (1), the similar analysis was performed in several
$t$-intervals: $[0,0.05]$, $[0.05,0.1]$, $[0.1,0.2]$, $[0.2,0.3]$,
$[0.3,0.5]$, $[0.5,0.7]$ $(GeV/c)^2$, and in each $t$-interval the
ratio of the $a_0$ event number to the total event number $R(a_0/tot)$
was determined for $[945,1035]$ MeV $\eta\pi^o$-mass interval.
This ratio was further approximated by an exponential function
of $t$. At the final stage of the analysis the PWA was performed
in $[945,1035]$ MeV mass interval independently in 20 $(MeV/c)^2$
bins of the transverse momentum squared. The $a_0(980)$ event numbers
in the $t$-bins were obtained then as the $S$-wave intensities 
corrected for corresponding $R(a_0/tot)$-factors, Fig.~3.

The obtained $t$-distribution was fitted by functions
$-N_1 \beta_1^2 t e^{\beta_1 t}$ and $N_2 \beta_2 e^{\beta_2 t}$
corresponding to $b_1$ and $\rho_2$ exchange respectively in
the $t$-channel of reaction (1) \cite{Achasov2}, as well as by a sum
of both functions. The fit by the first function only ($b_1$ exchange)
is quite unsatisfactory ($\chi^2/N_{DoF} = 116/33$) due to the peak at
$t\sim 0$ in the $t$-distribution, Fig.~3. The second function 
(solid line, $\chi^2/N_{DoF} = 27/33$) and the sum of 
both functions (dashed line, $\chi^2/N_{DoF} = 11/31$) fit the
measured differential distribution equally well, see Fig.~3.
\begin{figure*}[tbh]
\centerline{\epsfig{file=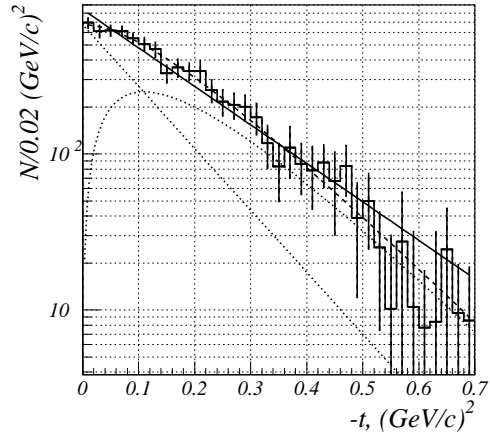,height=2.8in,width=2.8in}}
\vspace{5pt}
\caption{ The measured $t$-distribution of the $a_0(980)$-mesons produced
in reaction (1). The solid (dashed) line shows the fit of the distribution
by function corresponding to the $\rho_2$ ($\rho_2$ together with $b_1$) 
exchange in the $t$-channel of reaction (1). Dotted lines show the individual
$\rho_2$ and $b_1$ contributions in the case of the joint ~$\rho_2$~\&~$b_1$ 
fit.}
\end{figure*}
We conclude therefore that the $\rho_2$ exchange is quite necessary for
description of the differential cross section of the $a_0(980)$-meson
production in reaction (1) in agreement with 
paper \cite{Achasov2}.

\vspace*{-0.4cm}
\section*{Acknowledgements}
The authors would like to thank N.N.~Achasov for useful discussions.
One of the authors (S.A.S) gratefully acknowledges the financial 
support of the Local Organizing Committee of HADRON'97 Conference.




\begin{thebibliography}{99}
\bibitem{a0}
Jaffe~R.L., {\it Phys. Rev.} {\bf D15}, 267, 281 (1977);
~Montanet~L., {\it Rep.Prog.Phys.}\ {\bf 46}, 337 (1983);
~Close~F.E., {\it Rep.Prog.Phys.}\ {\bf 51}, 833 (1988);
~Achasov~N.N., {\it et al.}, {\it Usp.Fiz.Nauk}\ {\bf 142}, 361 (1984),
{\it Usp.Fiz.Nauk}\ {\bf 161}, 53 (1991).
\bibitem{etapi}
Prokoshkin~Yu.D. and Sadovsky~S.A., 
{\it Phys. Atom. Nucl.}\ {\bf 58}, 853 (1995).
\bibitem{a2}
Apel~W.D. {\it et al.},  {\it Nucl. Phys.}\ {\bf B193}, 269 (1981);
{\it Yad.Phys.}\ {\bf 41}, 126 (1985).
\bibitem{eff}
Sadovsky~S.A. {\it et al.}, {\it Proc. XXVI Intern.
Conf. HEP,}\ Dallas, 1791 (1992).
\bibitem{Tuan-Achasov}
Achasov~N.N. {\it et al.}, {\it Yad.Phys.}\ {\bf 33}, 1337 (1981);\\
Tuan~S.F. {\it et al.}, {\it Phys.Lett.}\ {\bf B213}, 537 (1988).
\bibitem{Achasov2}
Achasov~N.N., Shestakov~G.N.,  
{\it Phys.Rev.D}\ {56}, 212, (1997), hep-ph/9610409.
\bibitem{a4}
Alde~D. {\it et al.}, {\it Phys. Atom. Nucl.}\ {\bf 59}, 1027 (1996).
\bibitem{s8}
Sadovsky~S.A., Preprint IHEP 91-75, Protvino, (1991).
\bibitem{Flate}
Flatte~S., {\it Phys. Lett.}\ {\bf 63B}, 224 (1976).
\bibitem{Achasov3}
Achasov~N.N. {\it et al.},  
{\it Phys.Rev.D}\ {56}, 203, (1997), hep-ph/9605245.
\end{thebibliography}
\end{document}